\documentstyle[12pt,epsfig]{article}
\title{ Search for SUSY at LHC in Jets + $E^{miss}_T$ final states   
for the case of nonuniversal gaugino masses}
\author{\large S.I.~Bityukov$^1$ and  N.V.~Krasnikov  \\[3mm]
\em Institute for Nuclear Research RAS, \\
\em Moscow, 117312, Russia  }
\date{}
\begin{document}
\maketitle
\begin{abstract}
We investigate squark and gluino  pair production
at LHC (CMS) with subsequent decays into quarks and LSP for the case 
of nonuniversal gaugino masses. Visibility of signal by an excess
over SM background in $(n \geq 2)jets + E^{miss}_T$ events
depends rather strongly on the relation between LSP, gluino and 
squark masses and it decreases with the increase of LSP mass. 
For relatively heavy LSP mass closed to squark or gluino masses and 
for $(m_{\tilde{q}}, m_{\tilde{g}}) \geq 1.5$ TeV signal 
is too small to be observable. For the case when only third generation 
squarks and LSP are relatively light signal is not visible without 
b-tagging.

\end{abstract}
\vspace{1cm}
\bigskip

\noindent
\rule{3cm}{0.5pt}\\
$^1$~~Institute for High Energy Physics, Protvino, Russia

\section{Introduction}

One of the LHC supergoals is the discovery of the supersymmetry. 
In particular, it is very important to investigate a possibility
to discover strongly interacting superparticles (squarks and gluino). 
In ref.\cite{1} (see, also references \cite{2}) the LHC squark 
and gluino  
discovery potential
has been investigated within the minimal SUGRA-MSSM framework 
\cite{3} where all 
sparticle masses are determined mainly by two parameters: $m_0$ (common
squark and slepton mass at GUT scale) and $m_{1 \over 2}$ (common
gaugino mass at GUT scale). 
The signature used for the search for squarks and gluino  at LHC is 
$(n \geq 0)$ leptons + $(n \geq 2)jets + E^{miss}_T$ events. 
The conclusion of ref.~\cite{1}  is that LHC is able 
to detect squarks and gluino  with masses up to 
(2 - 2.5) TeV.  

In this paper we investigate the squark and gluino  discovery potential 
of LHC for the case of nonuniversal gaugino masses. Despite
the simplicity of the  SUGRA-MSSM framework it is a very particular
model. The mass formulae for sparticles in  SUGRA-MSSM model are derived
under the assumption that at GUT scale ($M_{GUT} \approx 2 \cdot 10^{16}$~GeV) 
soft supersymmetry breaking terms are universal. However, in general,
we can expect that real sparticle masses can differ in a drastic way 
from sparticle masses pattern of SUGRA-MSSM model due to many reasons,
see for instance refs.~\cite{4,5,6,7}.
Therefore, it is more appropriate to investigate the LHC SUSY discovery 
potential in a model-independent way \footnote{The early version of this 
study has been published in ref.~\cite{8}.}. 

The cross section 
for the production of strongly interacting superparticles
\begin{equation}
pp \rightarrow \tilde{g}\tilde{g}, \tilde{q}\tilde{g}, \tilde{q} \tilde{q}
\end{equation}
depends on gluino and squark masses.
Within SUGRA-MSSM  model the following approximate relations among sparticle 
masses take place:
\begin{equation}
m^2_{\tilde{q}} \approx m^2_0 + 6 m^2_{1 \over 2},
\end{equation}
\begin{equation}
m_{\tilde{\chi}^0_1} \approx 0.45 m_{1 \over 2},
\end{equation}
\begin{equation}
m_{\tilde{\chi}^0_2} \approx m_{\tilde{\chi}^{\pm}_1} \approx 
2m_{\tilde{\chi}^0_1},
\end{equation}
\begin{equation}
m_{\tilde{g}} \approx 2.5m_{1 \over 2}
\end{equation}
The decays of squarks and gluino depend on the relation among squark and 
gluino masses. For $m_{\tilde{q}} > m_{\tilde{g}}$ squarks decay mainly 
into gluino and quarks

\begin{itemize}

\item
$\tilde{q} \rightarrow \tilde{g} q$

\end{itemize}  
and gluino decays mainly into quark-antiquark pair and gaugino

\begin{itemize}

\item
$\tilde{g} \rightarrow q \bar{q} \tilde{\chi}^0_i$

\end{itemize}
\begin{itemize}

\item
$\tilde{g} \rightarrow q \bar{q}^{'} \tilde{\chi}^{\pm}_{1}$

\end{itemize}

For $m_{\tilde{q}} < m_{\tilde{g}}$ gluino decays mainly into squarks and 
quarks 

\begin{itemize}

\item

$\tilde{g} \rightarrow \bar{q} \tilde{q}, q \bar{\tilde{q}}$

\end{itemize}

whereas squarks decay mainly into quarks and gaugino

\begin{itemize}

\item
$ \tilde{q} \rightarrow   q \tilde{\chi^0_{i}}$

\end{itemize}
\begin{itemize}
\item
$\tilde{q} \rightarrow q^{'} \tilde{\chi}^{\pm}_1$

\end{itemize} 

The lightest chargino $\tilde \chi^{\pm}_1$ has several leptonic decay modes
giving a lepton and missing energy:

three-body decay

\begin{itemize}

\item 
$\tilde \chi^{\pm}_1 \longrightarrow \tilde \chi^0_1  + l^{\pm} + \nu$,

\end{itemize}

two-body decays

\begin{itemize}

\item
$\tilde \chi^{\pm}_1  \longrightarrow  \tilde l^{\pm}_{L,R} + \nu$,

\hspace{16mm}  $\hookrightarrow \tilde \chi^0_1 + l^{\pm}$

\item
$\tilde \chi^{\pm}_1 \longrightarrow \tilde \nu_L + l^{\pm}$,

\hspace{16mm} $ \hookrightarrow \tilde \chi^0_1 + \nu$

\item
$\tilde \chi^{\pm}_1 \longrightarrow \tilde \chi^0_1 + W^{\pm}$.

\hspace{26mm} $ \hookrightarrow l^{\pm} + \nu$

\end{itemize}

Leptonic decays of $\tilde \chi^0_2$ give two leptons and missing 
energy:

three-body decays

\begin{itemize}
\item 
$\tilde \chi^0_2 \longrightarrow \tilde \chi^0_1 + l^+ l^-$,

\item
$\tilde \chi^0_2 \longrightarrow \tilde \chi^{\pm}_1 + l^{\mp} + \nu$,

\hspace{16mm} $ \hookrightarrow \tilde \chi^0_1 + l^{\pm} + \nu$

\end{itemize}

two-body decay

\begin{itemize}
\item
$\tilde \chi^0_2 \longrightarrow \tilde l^{\pm}_{L,R} + l^{\mp}$.

\hspace{16mm} $ \hookrightarrow \tilde \chi^0_1 + l^{\pm}$

\end{itemize}

As a result of chargino and second neutralino leptonic decays besides 
classical signature

\begin{itemize}
\item
$(n \geq 2)$ jets plus $E^{miss}_{T}$
\end{itemize}
signatures

\begin{itemize}
\item
$(k \geq 1)$ leptons plus $(n \geq 2)$ jets plus $E^{miss}_{T}$
\end{itemize}
with leptons and jets in final state arise.
As mentioned above, these signatures have been used in ref.~\cite{1}
for investigation of LHC(CMS) squark and gluino discovery 
potential within SUGRA-MSSM  model, in which  gaugino masses 
$m_{\tilde \chi^0_1}$, $m_{\tilde \chi^0_2}$ are
determined mainly by a common gaugino mass $m_{1 \over 2}$. 

In our  
study we consider the general case when the relation between 
$m_{\tilde \chi^{0}_1}$ and $m_{\tilde{g}}$ is arbitrary. We investigate 
the detection supersymmetry using classical signature 
$(n \geq 2)$ jets plus $E^{miss}_{T}$. Signatures with several leptons 
in final state are more model dependent and besides classical 
signature leads to the highest discovery potential within 
SUGRA-MSSM  model.
We find that LHC squarks and gluino  discovery potential depends rather 
strongly on the relation between $\tilde \chi^0_1$, $\tilde g$ 
and $\tilde{q}$ masses and it decreases with the increase of the LSP mass.
 
\section{Simulation of detector response}

Our simulations are made at the particle level with parametrized
detector responses based on a detailed detector simulation. 
To be concrete our estimates have been made for the CMS(Compact Muon 
Solenoid)  detector. The CMS detector simulation 
program CMSJET~\cite{9} is used.
The main aspects of the CMSJET relevant to our study are the 
following.

\begin{itemize}
\item
Charged particles are tracked in a 4 T magnetic field. 90 percent 
reconstruction  efficiency per charged track with $p_T > 1$ GeV within 
$|\eta| <2.5$ is assumed. 

\end{itemize}

\begin{itemize}
\item
The geometrical acceptances for $\mu$ and $e$ are $|\eta| <2.4$ and 2.5, 
respectively. The lepton number is smeared according to parametrizations 
obtained from full GEANT simulations. For a 10 GeV lepton the momentum 
resolution $\Delta p_T/p_T$ is better than one percent over the full $\eta$ 
coverage. For a 100 GeV lepton the resolution becomes $\sim (1 - 5) \cdot 
10^{-2}$ depending on $\eta$. We have assumed a 90 percent triggering 
plus reconstruction efficiency per lepton within the geometrical 
acceptance of the CMS detector.  

\end{itemize} 

\begin{itemize}
\item
The electromagnetic calorimeter of CMS extends up to $|\eta| = 2.61$. There 
is a pointing crack in the ECAL barrel/endcap transition region 
between $|\eta| = 1.478 - 1.566$ (6 ECAL crystals). The hadronic calorimeter 
covers $|\eta| <3$. The Very Forward calorimeter extends from $|\eta| <3$ 
to $|\eta| < 5$. Noise terms have been simulated with Gaussian distributions 
and zero suppression cuts have been applied. 
\end{itemize}

\begin{itemize}
\item
$e/\gamma$ and hadron shower development are taken into account by 
parametrization of the lateral and longitudinal profiles of showers. The 
starting point of a shower is fluctuated according to an exponential 
law.

\end{itemize}

\begin{itemize}
\item
For jet reconstruction we have used a slightly modified UA1 Jet Finding 
Algorithm, with a cone size of $\Delta R = 0.8$ and 25 GeV transverse 
energy threshold on jets. 

\end{itemize}

\section{Backgrounds. SUSY kinematics}
  
All SUSY processes with full particle spectrum, couplings,
production cross section and decays are generated with ISAJET~7.32,
ISASUSY~\cite{10}. The Standard Model backgrounds are also generated 
with ISAJET~7.32. 

The following SM processes give the main contribution to the background:

\noindent
$WZ,~ZZ,~t \bar t,~Wtb,~Zb \bar b,~b \bar b$ and QCD $(2 \rightarrow 2)$ 
processes. 

As it has been mentioned above in our paper we consider only classical 
signature $(n \geq 2)$ jets plus $E^{miss}_{T}$ for squarks and gluino 
detection. We considered 3 different kinematical regions:

A. $m_{\tilde{g}} \gg m_{\tilde {q}}$

B. $m_{\tilde{q}} \gg m_{\tilde{g}}$

C. $m_{\tilde{q}} \sim m_{\tilde{g}}$, $m_{\tilde{q}} > m_{\tilde{g}}$

We also considered the case when all sparticles are heavy except the third 
generation sfermions and LSP \cite{12}.
For the case A   squarks $pp \rightarrow \tilde{q} \tilde{q}$ 
 production dominates at LHC. The squark  decays into quarks and LSP  
$\tilde{q} \rightarrow q \tilde{\chi}^0_1$ lead to the signature 
2 jets plus $E^{miss}_{T}$. For the case B  gluino pair production 
$pp \rightarrow \tilde{g} \tilde{g}$ dominates. The gluino 
decays $\tilde{g} \rightarrow q \bar{q} \tilde{\chi}^0_1 $ lead 
to $(n \geq 3)$ jets plus $E^{miss}_{T}$ signature. For the case C 
both squarks and gluino are produced $pp \rightarrow \tilde{q}\tilde{q}, 
\tilde{g} \tilde{g}, \tilde{q}\tilde{g}$ at the similar rate. Their 
decays give both 2 jets and $(n \geq 3)$ jets events. 
We considered two types of cuts:

Cuts a. $(n \geq 2)$ jets with $E_{Tjet1} \geq E_{T1}$, $E_{Tjet2} \geq 
E_{T2}$, $E^{miss}_T \geq E_{T0}$.

Cuts b. $(n \geq 3)$ jets with $E_{Tjet1} \geq E_{T1}$, 
$E_{Tjet2} \geq E_{T2}$, $E_{Tjet3} \geq E_{T3}$, $E^{miss}_T \geq 
E_{T0}$. 
Cuts a and b are  appropriate for the investigation of the 
kinematical points A and B correspondingly, whereas for the point C 
both cuts a and b are useful. We have calculated SM backgrounds for 
different values $E_{Tjet1}$, $E_{Tjet2}$, $E_{Tjet3}$, $E_{T0}$  
of the cut parameters. Our results are presented in Tables 1-2.  

\begin{table}[h]
    \caption{Cuts a and the corresponding value of background 
events for $L=10^5pb^{-1}$ .}
    \label{tab:Tab.1}
    \begin{center}
\begin{tabular}{|r|r|r|r|r|}
\hline
\# of cut& $E_t^{miss}$ [GeV]& $E_{t1}$ [GeV]& $E_{t2}$ [GeV]& $N_{b}$ \\ 
\hline
 1 & 200  & 40    &    40 &   4995783\\
 2 & 200  & 100 & 100 &  3292494\\
 3 & 200  & 100 & 150 &  3097944\\
 4 & 200  &  50 & 100 &  4478452\\
 5 & 400  & 200 & 200 &  180868\\
 6 & 400  & 200 & 300 &  173889\\
 7 & 400  & 100 & 200 &  247991\\
 8 & 600  & 300 & 300 &    8992\\
 9 & 600  & 300 & 450 &    7771\\
10 & 600  & 150 & 300 &   17662\\
11 & 800  & 400 & 400 &    1120\\
12 & 800  & 400 & 600 &     963\\
13 & 800  & 200 & 400 &    2708\\
14 & 1000 & 500 & 500 &     229\\
15 & 1000 & 500 & 750 &     183\\
16 & 1000 & 250 & 500 &     616\\
17 & 1200 & 600 & 600 &      38\\
18 & 1200 & 600 & 900 &      28\\
19 & 1200 & 300 & 600 &     115\\
\hline
\end{tabular}
    \end{center}
\end{table}

\begin{table}[b]
    \caption{Cuts b and the corresponding value of background 
events for $L =10^5pb^{-1}$ .}
    \label{tab:Tab.2}
    \begin{center}
\begin{tabular}{|r|r|r|r|r|r|}
\hline
\# of cut& $E_t^{miss}$ [GeV] & $E_1$ [GeV]& $E_2$ [GeV]& $E_3$ [GeV]& $N_b$ \\ 
\hline
 1  &  200 &  40 &  40 &  40 &  2953667\\
 2  &  200 & 100 & 125 & 150 &   957089\\
 3  &  200 & 167 & 208 & 250 &   315594\\
 4  &  200 & 233 & 292 & 350 &   104932\\
 5  &  200 & 300 & 375 & 450 &    79970\\
 6  &  400 & 100 & 125 & 150 &   151076\\
 7  &  400 & 167 & 208 & 250 &    20392\\
 8  &  400 & 233 & 292 & 350 &     9025\\
 9  &  400 & 300 & 375 & 450 &     4113\\
 10 &  600 & 100 & 125 & 150 &     8774\\
 11 &  600 & 167 & 208 & 250 &     4547\\
 12 &  600 & 233 & 292 & 350 &     2599\\
 13 &  600 & 300 & 375 & 450 &     1701\\
 14 &  800 & 100 & 125 & 150 &     1693\\
 15 &  800 & 167 & 208 & 250 &      754\\
 16 &  800 & 233 & 292 & 350 &      372\\
 17 &  800 & 300 & 375 & 450 &      194\\
 18 & 1000 & 100 & 125 & 150 &      425\\
 19 & 1000 & 167 & 208 & 250 &      234\\
 20 & 1000 & 233 & 292 & 350 &      147\\
 21 & 1000 & 300 & 375 & 450 &       59\\
 22 & 1200 & 100 & 125 & 150 &       99\\
 23 & 1200 & 167 & 208 & 250 &       58\\
 24 & 1200 & 233 & 292 & 350 &       31\\
 25 & 1200 & 300 & 375 & 450 &       22\\
\hline
\end{tabular}
    \end{center}
\end{table}

In this paper we considered the case when all squarks have the same mass 
and $m_{\tilde{\chi}^0_2}, m_{\tilde{\chi}^{\pm}_1} > min(m_{\tilde{g}},  
m_{\tilde{q}})$. The last requirement leads to the suppression of the 
events with leptons, only classical signature with $(n \geq 2)$ jets 
plus $E^{miss}_T$ is essential. The shape of the squark and gluino 
differential decay width depends rather strongly on the relation 
among squark, gluino and LSP masses. We considered 
different values of squark and gluino masses.  We took LSP mass equal to 
$m_{\tilde{\chi}^0_1} = k \times min(m_{\tilde{g}}, m_{\tilde{q}})$ with 
$k = \frac{1}{6}, 0.5, 0.75, 0.9$. The  $k = \frac{1}{6}$ corresponds 
approximately to the  standard case with universal gaugino masses.

\section{Results}

The results of our calculations are presented in Tables 3-6 and in 
Figures 1-6. In
estimation of the LHC(CMS) gaugino discovery potential we have used 
the significance determined as 
$\displaystyle S_{12} = \sqrt(N_s + N_b) - \sqrt(N_b)$ 
which is appropriate for the estimation of discovery potential in 
the case of future experiments~\cite{11}. 
We also imposed additional requirement that the ratio 
of signal to background events has to be bigger than 0.5 ($S/B > 0.5$).
 For the comparison 
we also give the values of 
often used significance \cite{1} determined as 
$\displaystyle S = \frac{N_s}{\sqrt(N_s + N_b)}$ and $5\sigma$ discovery 
probability~\cite{11} for two values of background $N_{back}$.           
Here $N_s = \sigma_s \cdot L$ is the average number of signal events and 
$N_b = \sigma_b \cdot L$ is the average number of background events for a 
given total luminosity L. 

\begin{table}[h]
    \caption{The number of events for cut with maximal value
of ``significances'' $S_2$ and $S_{12}$ and the  
corresponding value of the $5\sigma$ discovery probability 
for $L=10^5pb^{-1}$. Case A.}
    \label{tab:Tab.3}
    \begin{center}
\begin{tabular}{|c|c|r|r|r|r|cc|}
\hline
$M_{\tilde q}$&  &       &        &     &   & probability & of discovery \\
($M_{\tilde g}$),&$M_{\chi_1^0}$,&$cut\#$&$Signal$&$S_2$&$S_{12}$& 
$N_{back}=N_b$ & $N_{back} = 2 \cdot N_{b}$ \\
 GeV& GeV &       &        &     &   & (see~Tab.I) &  \\
\hline
 2450 &  400 &  18  &   33. &  4.23 &  2.52     &   0.667 & 0.211 \\
 (3050) & 1200 &  17  &   21. &  2.73 &  1.52     &   0.055 & 0.006 \\
\hline
 2050 &  350 &  19  &  176. & 10.32 &  6.33     &   1.000 & 1.000 \\
 (3000) & 1025 &  19  &   63. &  4.72 &  2.62     &   0.684 & 0.198 \\
  & 1500 &      &       &       &          &$\frac{N_s}{N_b}<0.5$ & \\
  & 1950 &      &       &       &          &$\frac{N_s}{N_b}<0.5$ &  \\
\hline
 1550 &  260 &  16  & 1773. & 36.27 & 24.06     &   1.000 & 1.000 \\  
 (2000) &  770 &  13  & 1697. & 25.57 & 14.33     &   1.000 & 1.000 \\  
  &  1162 &     &       &       &          &$\frac{N_s}{N_b}<0.5$ & \\    
  &  1395 &     &       &       &          &$\frac{N_s}{N_b}<0.5$ &  \\
\hline
 1050 &   175 & 10  &18793. & 98.43 & 58.03     &   1.000 & 1.000 \\    
 (2000) &   525 & 10  &10954. & 64.75 & 36.26     &   1.000 & 1.000 \\
  &   788 & 19  &   74. &  5.38 &  3.02     &   0.900 & 0.417 \\  
  &   945 & 19  &   86. &  6.07 &  3.45     &   0.983 & 0.681 \\ 
\hline
 550 &    92 &  7  &180270.&275.47 & 156.43     &   1.000 & 1.000\\
 (2000) &   225 & 10  & 20210.&103.85 &  61.71     &   1.000 & 1.000\\ 
  &   412 & 10  &  9939.& 59.82 &  33.24     &   1.000 & 1.000\\ 
  &   495 & 13  &  2560.& 35.27 &  20.54     &   1.000 & 1.000\\    
\hline
\end{tabular}
    \end{center}
\end{table}

\begin{table}[h]
    \caption{The number of events for cut with maximal value
of ``significances'' $S_2$ and $S_{12}$ and the 
corresponding value of the $5\sigma$ discovery probability 
for $L = 10^5 pb^{-1}$ . Case B.}
    \label{tab:Tab.4}
    \begin{center}
\begin{tabular}{|c|c|r|r|r|r|cc|}
\hline
$M_{\tilde g}$&  &       &        &     &   & probability & of discovery \\
($M_{\tilde q}$),&$M_{\chi_1^0}$,&$cut\#$&$Signal$&$S_2$&$S_{12}$& 
$N_{back}=N_b$ & $N_{back} = 2 \cdot N_{b}$ \\
 GeV& GeV &       &        &     &   & (see~Tab.II) &  \\
\hline
 2000 &   350 & 24  &   18.& 2.57 &  1.43  &  0.031 & 0.004 \\    
 (2950) &  1000 &     &       &       &         &$\frac{N_s}{N_b}<0.5$  & \\
\hline
 1500 &   251 & 16  &   720.& 21.79 &  13.75  &  1.000 & 1.000 \\    
 (1950) &   750 & 15  &   456.& 13.11 &   7.32  &  1.000 & 1.000 \\  
  &  1125 &     &       &       &         &$\frac{N_s}{N_b}<0.5$  & \\
  &  1350 &     &       &       &         &$\frac{N_s}{N_b}<0.5$  &  \\
\hline
 1000 &   175 &  7  & 10970.& 61.94 &   34.29 &  1.000 & 1.000 \\    
 (1950) &   500 & 11  &  3050.& 34.99 &   19.73 &  1.000 & 1.000 \\ 
  &   750 & 14  &  1271.& 23.34 &   13.29 &  1.000 & 1.000 \\ 
  &   900 & 14  &  1100.& 20.81 &   11.70 &  1.000 & 1.000 \\ 
\hline
  500 &    94 &  2  &1164159.&530.81&  300.77 &  1.000 & 1.000 \\ 
 (1950) &   250 &  7  &  30710.&135.85&   83.26 &  1.000 & 1.000 \\   
\hline
\end{tabular}
    \end{center}
\end{table}

\begin{table}[h]
    \caption{The number of events for cut with maximal value
of ``significances'' $S_2$ and $S_{12}$ and the  
corresponding value of the $5\sigma$ discovery probability 
for $L=10^5pb^{-1}$. Case C, cuts a.}
    \label{tab:Tab.5}
    \begin{center}
\begin{tabular}{|c|c|r|r|r|r|cc|}
\hline
$M_{\tilde q}$&  &       &        &     &   & probability & of discovery \\
($M_{\tilde g}$),&$M_{\chi_1^0}$,&$cut\#$&$Signal$&$S_2$&$S_{12}$& 
$N_{back}=N_b$ & $N_{back} = 2 \cdot N_{b}$ \\
 GeV& GeV &       &        &     &   & (see~Tab.I) &  \\
\hline
 2400 &  400 &  19  &   85. &  6.01 &  3.41  &   0.980 & 0.661 \\
 (2300)& 1150&      &       &       &          &$\frac{N_s}{N_b}<0.5$ & \\
  & 1700 &      &       &       &          &$\frac{N_s}{N_b}<0.5$ & \\
\hline
 2100 &  350 &  19  &  191. & 10.92 &  6.77     &   1.000 & 1.000 \\
 (2000) & 1000 &  19  &   63. &  4.72 &  2.62   &   0.684 & 0.198 \\
  & 1500 &      &       &       &          &$\frac{N_s}{N_b}<0.5$ & \\
\hline
\end{tabular}
    \end{center}
\end{table}

\begin{table}[h]
    \caption{The number of cut with maximal value
of ``significances'' $S_2$ and $S_{12}$ and 
corresponding value of probability of discovery. Case C, cuts b.}
    \label{tab:Tab.6}
    \begin{center}
\begin{tabular}{|c|c|r|r|r|r|cc|}
\hline
$M_{\tilde g}$&  &       &        &     &   & probability & of discovery \\
($M_{\tilde q}$),&$M_{\chi_1^0}$,&$cut\#$&$Signal$&$S_2$&$S_{12}$& 
$N_{back}=N_b$ & $N_{back} = 2 \cdot N_{b}$ \\
 GeV& GeV &       &        &     &   & (see~Tab.2) &  \\
\hline
 2300 &   400 & 22  &   73.& 5.55 &  3.16  &  0.943 & 0.508 \\    
 (2400) & 1150 & 24  &   18.& 2.57 & 1.43  &  0.031 & 0.004 \\  
  & 1700 &     &       &       &         &$\frac{N_s}{N_b}<0.5$  & \\
\hline
 2000 &   350 & 16  & 283.& 11.06 &  6.30 &  1.000 & 1.000 \\    
 (2100) & 1000 & 22  & 51.& 4.15 &   2.29 &  0.446 & 0.088 \\ 
  & 1500 &     &       &       &         &$\frac{N_s}{N_b}<0.5$  & \\
\hline
 1500 &   251 & 14  & 2910.& 42.89 &  26.70  &  1.000 & 1.000 \\    
 (1550) & 750 & 14  & 1180.& 22.01 &  12.45  &  1.000 & 1.000 \\  
  & 1125 &     &       &       &         &$\frac{N_s}{N_b}<0.5$  & \\
  & 1350 &     &       &       &         &$\frac{N_s}{N_b}<0.5$  & \\
\hline
 1000 &   167 & 7  & 51260.& 191.50 & 124.88  &  1.000 & 1.000 \\    
 (1050) & 500 & 7  & 24240.& 114.74 &  68.46  &  1.000 & 1.000 \\  
        & 750 & 10 &  5700.&  47.38 &  26.64  &  1.000 & 1.000 \\  
        & 900 & 14 &  1460.&  26.00 &  15.00  &  1.000 & 1.000 \\  
\hline
  500  &  84 &  1  &4330000.&1604.4& 980.20 &  1.000 & 1.000 \\ 
 (550) & 250 &  1  &3456000.&1365.1& 813.11 &  1.000 & 1.000 \\   
       & 375 &  1  &1794000.& 823.3& 460.28 &  1.000 & 1.000 \\  
       & 450 &  6  & 108528.& 213.0& 120.83 &  1.000 & 1.000 \\  
\hline
\end{tabular}
    \end{center}
\end{table}

As it follows from our results for fixed 
values of squark and gluino masses the visibility of signal decreases with 
the increase of the LSP mass. This fact has trivial explanation. Indeed, in the 
rest frame of squark or gluino the jets spectrum becomes more soft with the 
increase of LSP mass. Besides in parton model pair produced squarks and 
gluino are produced with total transverse momentum closed to zero. For high 
LSP masses partial cancellation of missing transverse momenta from two 
LSP particles takes place. The fact that with rise of LSP mass the 
$E^{miss}_T$ spectrum becomes more soft is explicitly seen in Figures 1-3.

\begin{figure}[ht]
\epsfig{file=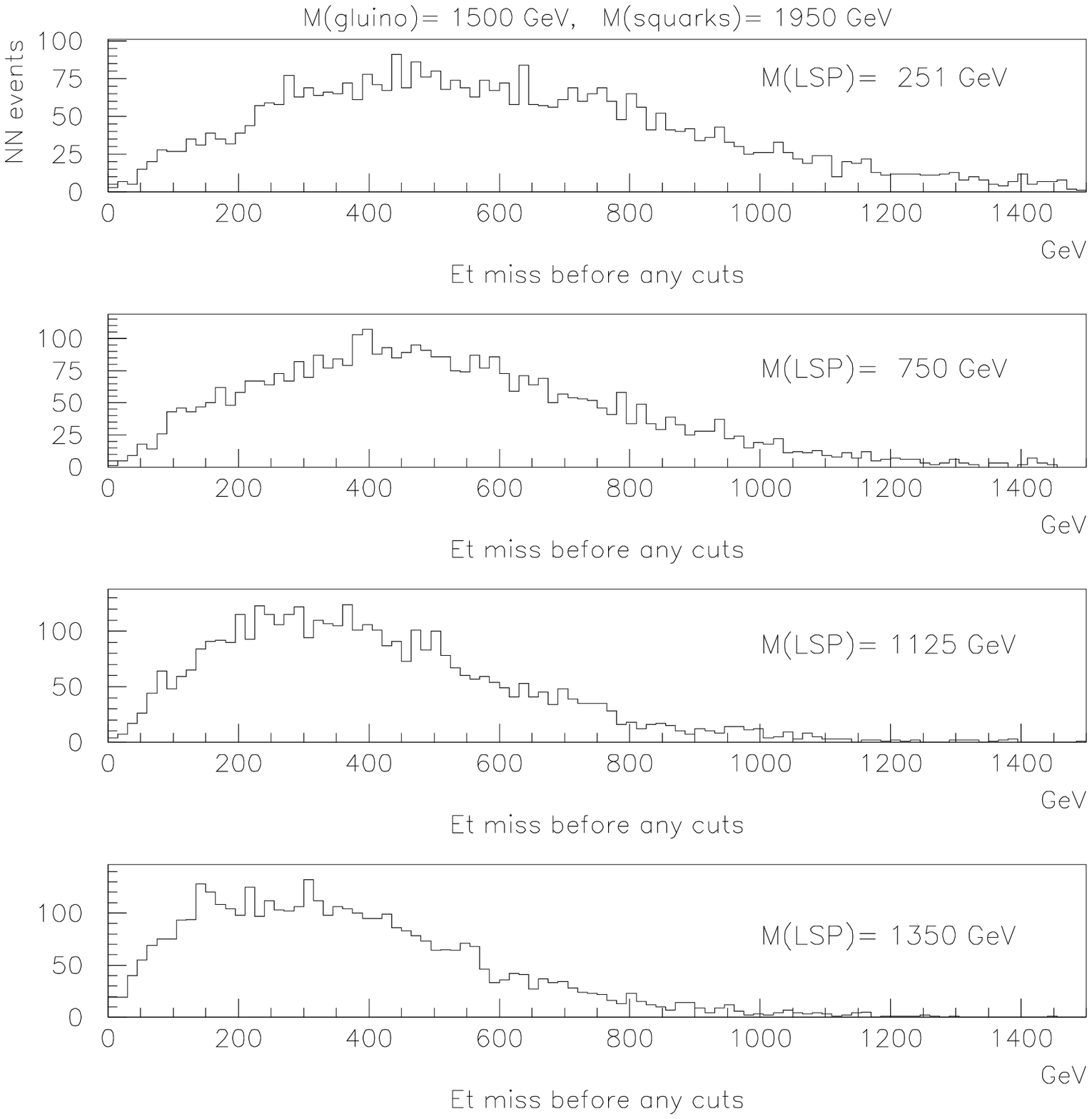,width=15cm}
\caption{The Et miss distribution for different LSP masses 
($M_{\bar g} = 1500~GeV,~M_{\bar q} = 1950~GeV$). Case B.}
\label{fig.1}
\end{figure}

\begin{figure}[ht]
\epsfig{file=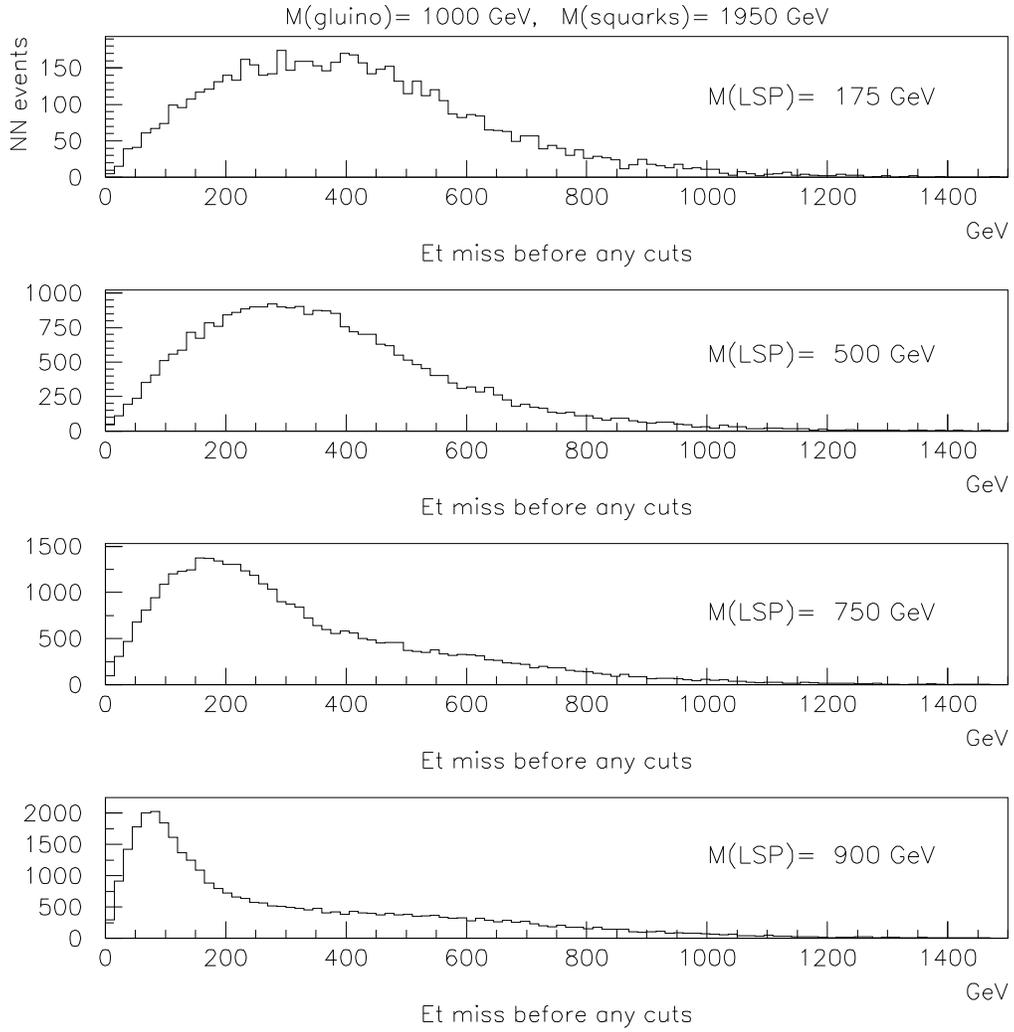,width=15cm}
\caption{The Et miss distribution for different LSP masses
($M_{\bar g} = 1000~GeV,~M_{\bar q} = 1950~GeV$). Case B.}
\label{fig.2}
\end{figure}

\begin{figure}[ht]
\epsfig{file=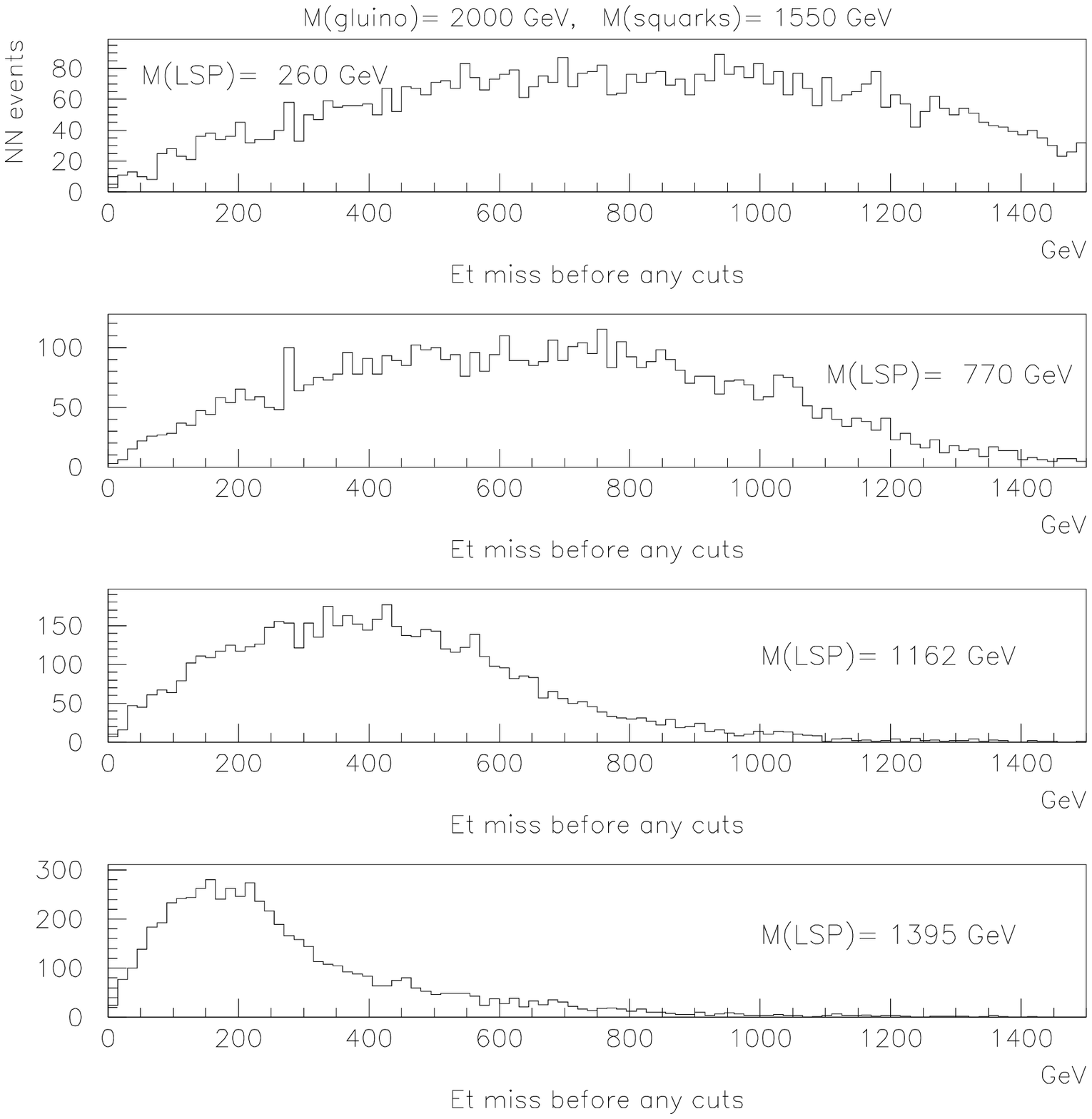,width=15cm}
\caption{The Et miss distribution for different LSP masses
($M_{\bar g} = 2000~GeV,~M_{\bar q} = 1550~GeV$). Case A.}
\label{fig.3}
\end{figure}

Figures 4-6 demonstrate that within well defined cut the number of signal 
events decreases with increase of the LSP masses that complicates 
supersymmetry detection for  LSP masses closed to gluino or squark 
masses. For the case when all sparticles are heavy except third squark 
generation and LSP we have found that signal is too small to 
be observable for all values of third generation squark masses and LSP mass. 
In this cass possible b-tagging will be able to suppress the background an 
make signal observable~\footnote{S.I.Bityukov and N.V.Krasnikov, the work 
is in progress.}.    

\begin{figure}[ht]
\epsfig{file=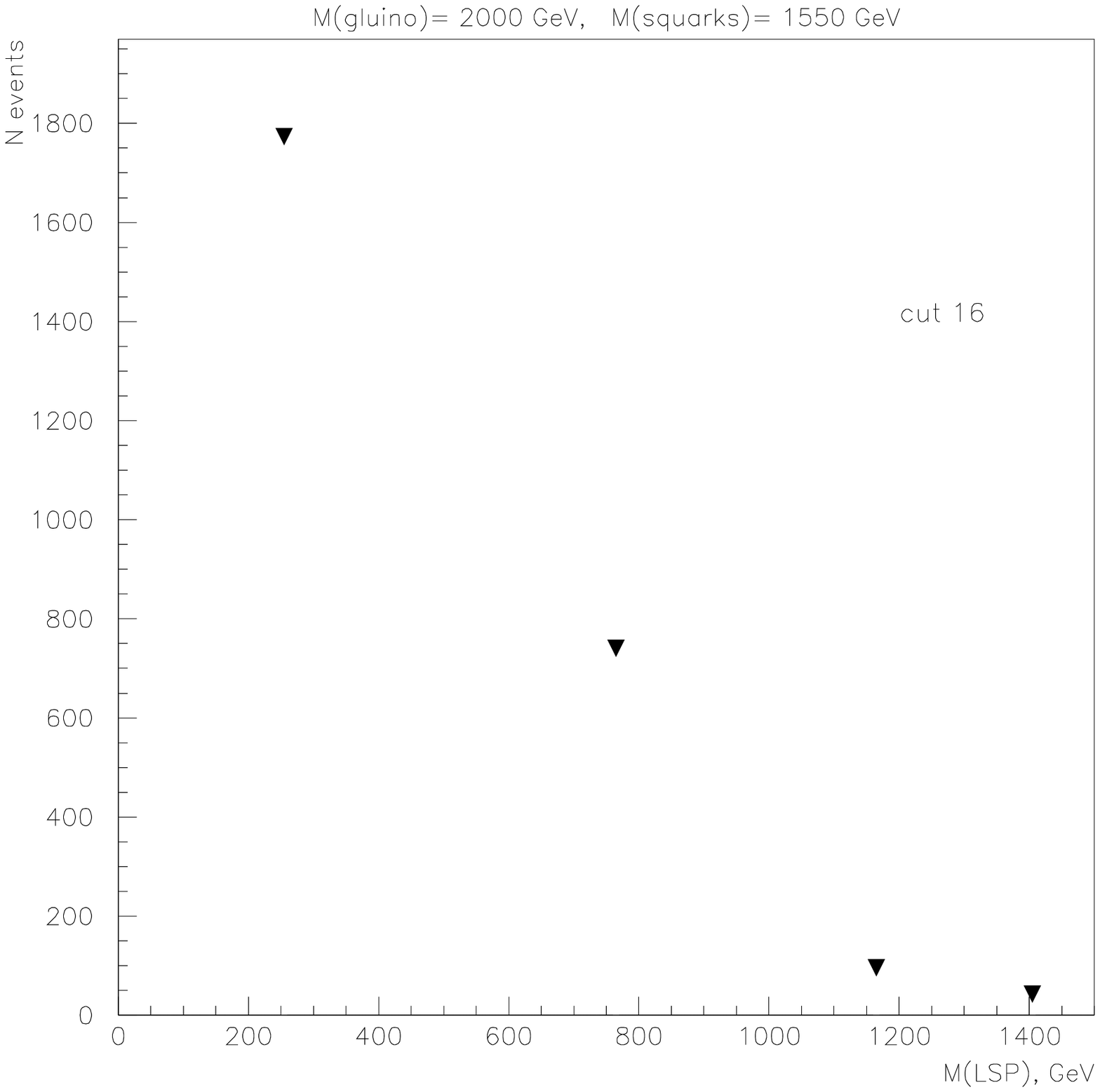,width=15cm}
\caption{The dependence of the signal events on the LSP mass. Case A, cut 16a.}
\label{fig.4}
\end{figure}

\begin{figure}[ht]
\epsfig{file=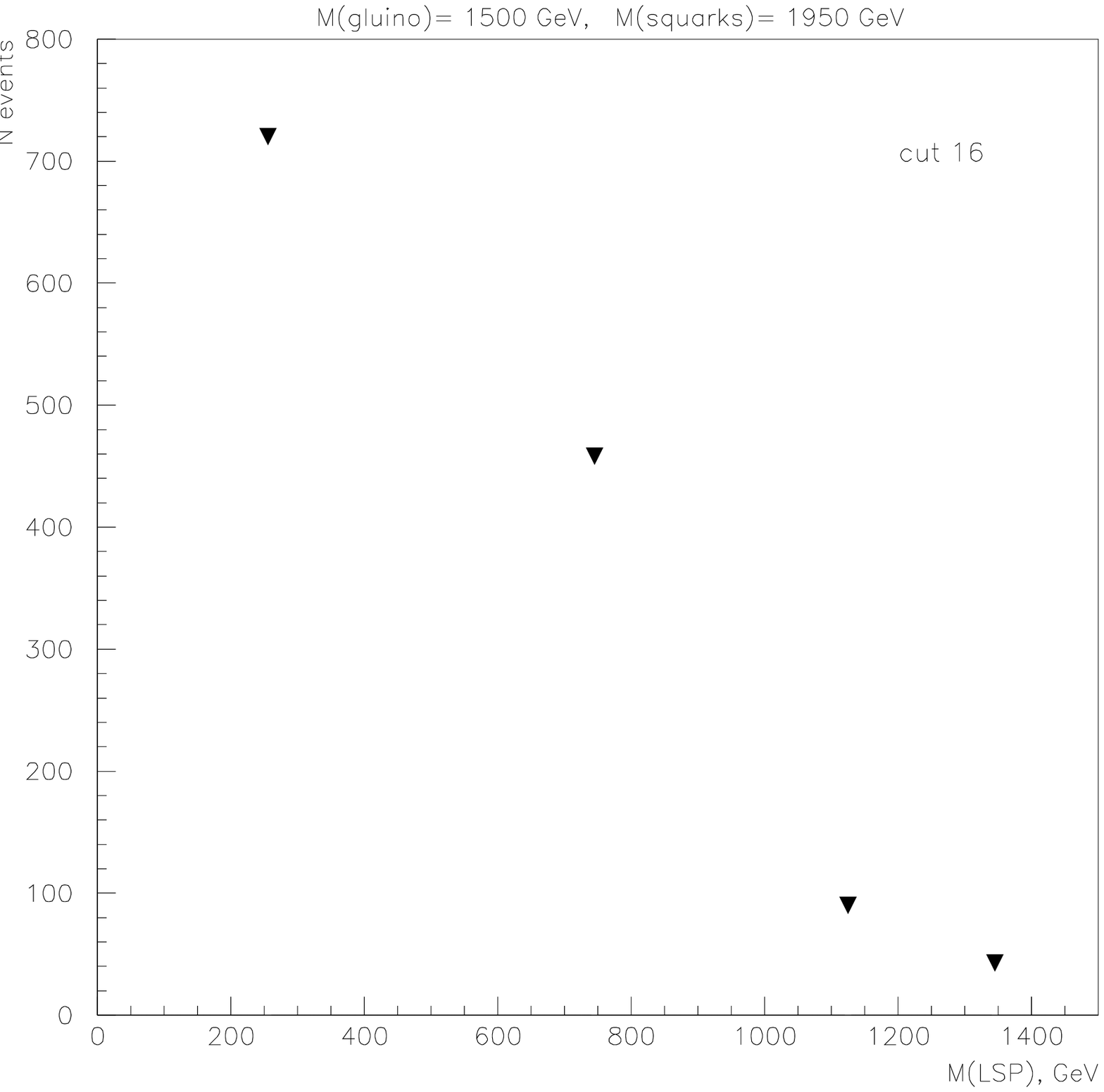,width=15cm}
\caption{The dependence of the signal events on the LSP mass. Case B, cut 16b.}
\label{fig.5}
\end{figure}

\begin{figure}[ht]
\epsfig{file=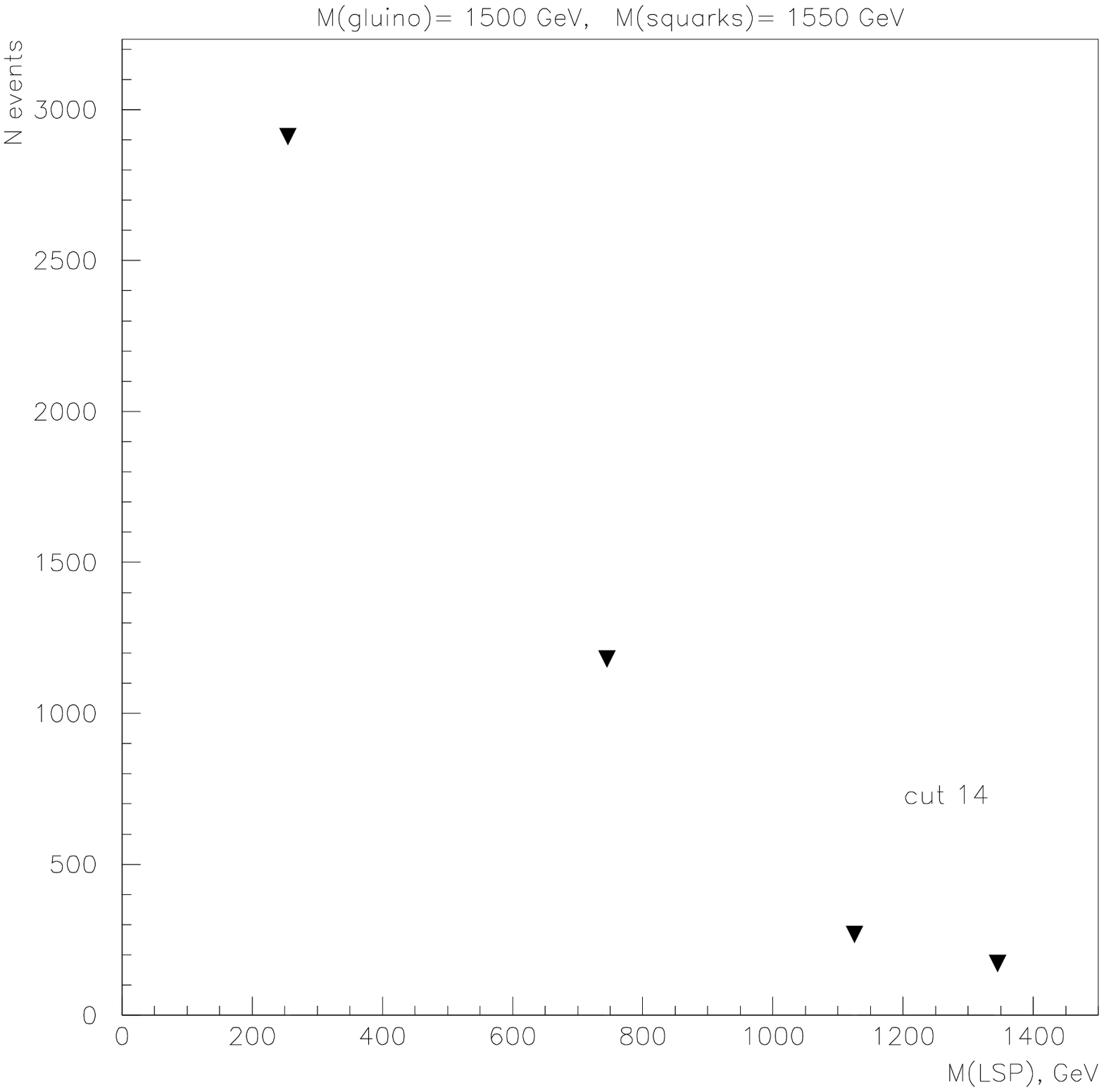,width=15cm}
\caption{The dependence of the signal events on the LSP mass. Case C, cut 14b.}
\label{fig.6}
\end{figure}

\section{Conclusion}

In this paper we have presented the results of the calculations for 
squark and gluino  pair production at LHC (CMS) with their 
subsequent decays into jets for the case of nonuniversal gaugino masses.
We have found that the visibility of signal by an excess over SM background
in $(n \geq 2)~jets + E^{miss}_T$ events depends rather 
strongly on the relation 
between LSP mass $\tilde \chi^0_1$ and  $\tilde{q}, \tilde{g}$ masses. 
The visibility of the signal for 
fixed values of squark and gluino masses decreases with the 
increase of the LSP mass.  
For relatively heavy LSP mass closed to gluino or squark masses and 
for $(m_{\tilde{g}}, m_{\tilde{q}}) \geq 1.5~TeV$  signal is too small 
to be observable.

\begin{center}
 {\large \bf Acknowledgments}
\end{center}

\par
We are  indebted to 
the participants of Daniel Denegri seninar on physics 
simulations at LHC for useful comments. 
This work has been supported by RFFI grant 99-02-16956.

%\bigskip
%\newpage

\end{document}